\documentstyle[preprint,revtex]{aps}

\begin{document}
\draft

\begin{title}
{\bf  The (001) surface of $\alpha$-Ga is covered with GaIII}
\end{title}

\author{M. Bernasconi$^{1}$, G. L. Chiarotti$^{1}$, E. Tosatti$^{1,2}$}

\begin{instit}
$^{1}$International School for Advanced Studies (SISSA),
Via Beirut 2,
I-34014 Trieste, Italy

$^{2}$ International Centre for Theoretical Physics (ICTP),
P.O. Box 586, I-34014 Trieste, Italy
\end{instit}

\begin{abstract}
We propose, based on ab initio calculations, that  the ground state
(001) surface of $\alpha$-Ga should be covered
 by two layers of  GaIII, a denser phase which is
stable in the bulk only at high pressure and temperature.
 We discuss how this novel prediction
can explain recent STM observations, including  anomalous thermal stability
 of this surface.
\end{abstract}

\nonum

\narrowtext

\section{}
When two or more phases compete for the ground state of a bulk system
there is a chance to find strange phenomena at the surface of the lowest energy
phase. One such scenario is the formation
of a thin film of one of the losing structures
at the surface of the  winning  bulk phase. There are known examples
of this behaviour. In surface melting when the temperature is still below the
bulk melting point, most  crystal surfaces ``self-wet'' with a thin
liquid film \cite{melting}. Another example is the valence transition
at the surface
of Sm
metal, which has valence $3^+$ in the bulk but becomes $2^+$ at the surface,
in spite of 0.26 eV/atom cohesive energy difference between  bulk
Sm$^{3+}$  and  Sm$^{2+}$ \cite{Sm}.
These ``self wetting'' phenomena can  be rationalized on the basis of a common
ingredient: the emergent surface phase has a sufficiently lower surface free
energy to make it worthwhile paying for the film
plus interface free energy costs.
The physical mechanisms underlying  the
surface free energy gain of the emergent phase may
however have very different origin.
In the case of surface melting, the surface free energy of the liquid is
lower than that of the solid, mostly due to entropy
\cite{melting}.
In metallic Sm the surface energy of the
Sm$^{2+}$ state is lower,  due to  closer
similarity to the divalent  ground state of the free atom \cite{Sm}.
 These surface
 energy gains
are usually quite small,
and the closeness of the bulk free energies
is thus a necessary condition for the self wetting to appear.

It is known both experimentally \cite{Bosio} and theoretically \cite{gong}
 that solid Ga presents
a rather complicated phase diagram, with many stable and metastable
phases closely competing for the ground state. The stable phase at low
temperature and ambient pressure is $\alpha$-Ga, which presents a remarkable
coexistence of metallic and covalent character. The energetically closest
phases are  GaII \cite{Bosio,notastru},
GaIII \cite{Bosio} and $\beta$-Ga \cite{bosiobeta},
 which are denser, and fully metallic.

In this letter, based on accurate ab-initio total energy calculations, we
propose that a thin film of the GaIII phase should  cover the
(001) surface of $\alpha$-Ga in its ground state, therefore providing a
novel realization of the self-wetting scenario.

The $\alpha$-Ga lattice is base-centered orthorhombic with 8 atoms in the
conventional unit cell \cite{wycoff} (inset of fig. 1).
 Each atom has  one neighbour only in the first coordination
cell at $2.44 \rm\AA$, with six further neighbours  within   $0.39 \rm\AA$.
The structure can be seen
as the stacking along the [001] direction of strongly buckled metallic planes
connected by short covalent bonds \cite{gong}.
The structure of the $\alpha$-Ga(001)  surface  has been recently
investigated
by  Scanning Tunnel Microscopy (STM) \cite{zugerprb,zugertesi}, where it
 appears to be unusually stable: no step diffusion or
other type of surface mobility
was detected  up to the bulk melting point ($T_m=303 K$). Even more
surprisingly,
at $T_m$ the Ga crystal begin to melt from {\sl inside}  the sample,
with the surface still appearing flat, and atomically ordered \cite{zugertesi}!
Besides, this surface raises additional questions.
In principle, in fact, the ideal $\alpha$-Ga (001) surface
 can be formed in two ways,
 by cutting the crystal   at a)
a plane
that separates  dimer layers, without cutting ``covalent'' bonds
(surface A) (fig.1a), or at b)  a plane that cuts the dimer covalent
bonds
(surface B)
 (fig.1b).
The top-view geometries for the ideal surfaces A and B are the same,
 and can be described
by a nearly square lattice
with
 two atoms per surface cell,
 forming chains along the [100] direction,
 with coordinates (in lattice units)
  (0,0,0) and (0.5,0.34,0).
The STM
map, shown in fig.2a,  clearly shows this chain structure
 ( a small dimerization of the chain was also
inferred   \cite{zugerprb}).
Now,  if both configurations A and B
were simultaneously  realized, for example in adjacent domains, then
steps with a height of $c/4 \simeq 1.9 \rm\AA$ should be present; conversely if
one of the two surfaces had a much lower  energy,
 the smallest step height expected is
 $c/2 \simeq 3.8 \rm\AA$. The observed step-height distribution
\cite{zugerprb}
supports the latter scenario, with just a single  step of $3.8 \rm\AA$
height. However, this still does not distinguish between possibilities A,  B,
 or others.

We have studied both surfaces A and B  by means of total energy
calculation within Local Density Approximation \cite{perdewZ}. We have used
first-principle norm conserving pseudopotential of the Kleinman Bylander
form \cite{Stumpf}.  The surfaces have been modeled by periodic slabs, 10 to 14
layers thick, separated by four vacuum layers.
Kohn-Sham orbitals were expanded in plane waves up to
 14 Ry
energy cutoff.
Convergency of  k-sums have been tested up to 49 k points in the
irreducible
 ideal surface BZ
(ISBZ); gaussian spreading with variance from 20 to 5 mRy has been used.
 The lattice parameters were obtained by an independent bulk
calculation \cite{kpoint}.
Guided by the calculated Hellmann-Feynman forces we let
the atoms in the slab relax to their lowest energy positions, with residual
forces less than $1.5 \, 10^{-3} Ry/a_o$.

 Figures 1a) and 1b)  show the electronic (pseudo) charge density
 on a (100)
plane, for fully relaxed configurations of surfaces A and B respectively.
 Surface A  has  energy
$\sigma =$ 70 mRy/atom  when unrelaxed,
and 57 mRy/atom when fully relaxed.
For the unrelaxed and relaxed B surfaces we find  $\sigma =$  59
and 57 mRy/atom respectively \cite{coor}.
 Surface energies were obtained by subtracting from the
slab total energy  the bulk energy for the same number
of atoms, calculated with a $\rm k$-points sampling as
 close as possible to that used in the slab
calculation, and are
 converged within 1 mRy/atom.
 These
values for the surface energies  appear to be
high when compared with  the $T=0 K$ extrapolation
of the experimental result $\sigma_{expt.}=$ 41 mRy/atom \cite{sigmaexp}.
The two relaxed surfaces A and B end up having the same
 energy, which does not explain  the  STM
step-height distribution.
 Moreover,
the strong corrugation of  charge density in surface A (fig.1a),
looks unreasonable, and indirectly suggests the existence of a  different
ground state. Also, the chain structure in surface A is different from
 the STM image.
 Surface B is instead far more compact, and its chain structure is very
similar to the STM image,
but it has one unsaturated dangling bond
per atom, which again suggests instability.
Guided by this reasoning we rearranged drastically surface A, by pushing
the outermost atom down to fill the hole underneath. This is equivalent to
adsorbing a plane of adatoms onto surface B, therefore removing the dangling
bonds.
We find  a  dramatic lowering of $\sigma$ down to 47 mRy/atom, which now
 compares much better with the experimental value.
In fig. 2 we compare the STM image calculated for the new optimal
surface configuration  (hereafter referred to as C)
with the experiment \cite{zugerprb}.
The characteristic chain present in the  data is very well
reproduced by our model.
  Also the non-spherical shape of the
experimental current spots, with a
 flat region at one side, compares well with our theoretical image.
Our results do not explain the apparent chain dimerization,
 but this effect probably involves energies below our
present resolution.

The charge density of surface C
on the (100) plane is plotted in fig.~3.
Remarkably, the structure of the two outermost surface layers
and their optimal charge density have become very reminiscent of those of
bulk GaIII (inset).
Bulk  GaIII is a tetragonally
distorted FCC structure, stable  at high pressure and
temperature \cite{Bosio}.
The chain structure at surface C is not present in the
geometry of  GaIII.
 However we have checked that, by forcing GaIII to
grow epitaxially on $\alpha$-Ga
(i.e in plane lattice parameters fixed to those of $\alpha$-Ga),
the FCC-like symmetry is unstable and the
chain structure appear also in bulk ``epitaxial'' GaIII. In actual fact, at the
surface of a slab of epitaxial GaIII the structure of the top
layer is identical  to that of our surface C.
We also find that the surface energy
 of epitaxial GaIII is 43 mRy/atom. This value is sufficiently
lower
than the surface energy of the ideal configurations ($\sigma_A = \sigma_B =$
57 mRy/atom) to
make it worth paying for the  $\alpha$-Ga/GaIII interface energy
 plus the
 difference in bulk energies of $\alpha$-Ga and epitaxial GaIII
($\Delta E = E_{bulk}$(epitax. GaIII) - $E_{bulk}$($\alpha$-Ga) which we have
calculated to be $\sim 5$
 mRy/atom) in the
 film covering the surface in
configuration C.
All these results support the prediction that
{\sl in the ground state the (001) surface of $\alpha$-Ga should be
 wetted by two
layers of GaIII epitaxially grown on $\alpha$-Ga}.

Since by adding a third GaIII layer one produces a step height roughly
one half  the experimental value, the wetting must be confined to the
first two layers up to $T_m$, in order to be consistent with  STM.
Indeed, by a separate calculation, we find that the surface energy of
 the relaxed configuration with {\sl three} layers of
GaIII is as high as  60 mRy/atom, similar to $\sigma_B$. These values,
 compared with
that of  surface C ($\sigma_C$ = 47 mRy/atom),
guarantee that the configurations with one and two, or with
two and three GaIII layers cannot
be simultaneously present, in agreement with the observed step
height distribution. Furthermore the surface energy for three GaIII layers is
much higher than the value obtained by adding to $\sigma_C$  the
energy difference $\Delta E \sim 5$ mRy/atom, required by the
 added GaIII plane. This suggests a strong
attraction between the surface and the $\alpha$-Ga/GaIII interface, which
should  also exclude the configurations with 4 or 5 planes of GaIII
(not investigated). Therefore the wetting of $\alpha$-Ga by epitaxial GaIII
is incomplete.

The surface-interface attraction produces a 3D atomic density in the
GaIII film roughly 8 $\%$ higher than the bulk equilibrium density of
epitaxial GaIII. If this  density increase were obtained by hydrostatic
 pressure it would cause
 GaIII
to melt about 100 K above the melting temperature of $\alpha$-Ga
  \cite{extratm}.
We believe that this explains the anomalous thermal stability
 of the $\alpha$-Ga(001)
 surface detected experimentally \cite{zugerprb,zugertesi}.

The presence of the GaIII film induces a large increase in the
surface-projected
electronic density of states at the Fermi level
with respect to the bulk density of states, characterized
 instead by pronounced pseudogap at $E_f$ \cite{gong,bulk}
(surface band structure and densities will be
published elsewhere \cite{bulk}). This ``metallization'' of the surface
 should show up in
photoemission measurements. The GaIII film should induce
 also detectable modifications
in the phonon spectra: the bulk phonon modes around 7 Thz  \cite{phonon},
 associated to the
stretching of the dimer covalent bond, should be absent
or strongly modified in the surface projected
density of states.  A more decisive test of our picture should of course
 come from
structural tools, such as ion scattering, dynamical LEED,
X-ray or atom diffraction, which should be able to discriminate
between A, B, and C structures. Static LEED is not sufficient, since the 2D
 space group is the same for A, B and C surfaces.

Finally, we may wonder whether this kind of solid-state
incomplete wetting is a peculiar feature  of
$\alpha$-Ga(001) surface or  could be envisaged  for other solids too.
In general the presence of half-filled dangling bonds
makes the surface energy of unreconstructed semiconductors
  higher than the
surface energy of metals with respect to their cohesive energy.
If a simple  reconstruction,
 still preserving the ``covalent bonding''
 is not feasible, then the wetting with a fully metallic phase
should in principle be favoured. This seems to be the case for
the $\alpha$-Ga(001), where the plausible structure of the ideal surface (B)
does retain covalent unsaturated dangling bonds.
Two other circumstances make the transformation possible on $\alpha$-Ga(001),
namely the
 closeness in energy between $\alpha$-Ga and ``epitaxial'' GaIII, as well as a
 strong interface attraction which appears to
limit the wetting film thickness to strictly  two atomic layers.

We are grateful to U. D\"{u}rig and O. Z\"{u}ger for discussions and
information, and to
S. Baroni and P. Giannozzi for providing us the DFT Fortran code.
This work has been supported by the Italian Consiglio
Nazionale delle Ricerche through
{\sl Progetto Finalizzato Sistemi Informatici e Calcolo Parallelo},
by INFM, and by
the European Office of the US Army.


\newpage

\figure{ a) Charge density of relaxed surface A plotted onto the (100) plane
passing through all atoms shown, $\sigma = 57$ mRy/atom; b)
same as in a) for surface B, $\sigma = 57$ mRy/atom.
 Inset: orthorhombic  unit cell of $\alpha$-Ga
\protect\cite{wycoff}.}

\figure{ a) Experimental STM image of $\alpha$-Ga(001) \protect\cite{zugerprb}.
b) Local
density of states $\rho({\bf r},\Delta E)$
calculated for surface C at  a plane  $z \simeq 2 \rm\AA$  above
surface atoms, and including states inside an energy window
$\Delta E = 0.5$ eV below the Fermi level.
 In the simplest approximation \protect\cite{tersoff} this plot
corresponds to the theoretical STM image.
States from 36 {\bf k} points at the boundary of the ISBZ have been
included. Note the close similarity of features in the two pictures.}

\figure{Charge density for our proposed optimal
structure for $\alpha$-Ga(001) (surface C).  $\sigma = 47$ mRy/atom.
The fully relaxed
interlayer distance
 for the outermost layers $d_{12}, d_{23}, d_{34}$ are  0.380,
0.347, and 0.525 respectively, in
unit of $a_0 = 4.377 \rm\AA$. The corresponding (x,y) in-plane coordinates
 (in lattice units)  of the two atoms per cell in
the four outermost planes are  (0,0) (0.5,-0.324); (0,0.505)
 (0.5,0.171); (0,0.002) (0.5,-0.326); (0,0.179) (0.5,-0.502) respectively.
 The two
outermost surface layers now mimic closely the bulk GaIII phase (inset).}


\begin{references}
\bibitem[6]{notastru} The structure named GaII in ref.
\protect\cite{gong} is more commonly referred to as
 GaIII in \protect\cite{Bosio}. We follow the latter in the present work.
\bibitem[18]{coor} Despite the stiffness of the covalent bond, the change
 in atomic coordination ( surface B is obtained by breaking only one stiff
covalent bond, while 4 metallic "bonds" are broken in surface A) makes
the unrelaxed surface A higher in energy than unrelaxed surface B.
\end{references}
\end{document}